\documentclass[10pt]{iopart}
\usepackage{graphicx}
\usepackage{stfloats}

\usepackage{amstext}
\usepackage{hyperref}
\usepackage{xcolor}

\begin{document}

\title[Experiments on intense ion beam formation with inhomogeneous electric field
]{Experiments on intense ion beam formation with inhomogeneous electric field
}

\author{S.~S.~Vybin, V.~A.~Skalyga, I.~V.~Izotov, S.~V.~Golubev, S.~V.~Razin, R.~A.~Shaposhnikov, M.~Yu.~Kazakov, A.~F.~Bokhanov, S.~P.~Shlepnev}

\address{Institute of Applied Physics of Russian Academy of Sciences, 46 Ulyanov Str.,
603950, Nizhny Novgorod, Russia}

\ead{vybinss@ipfran.ru}
\vspace{10pt}

\begin{abstract}
We report on a successful test of a new method for intense ion beam formation using an extraction with inhomogeneous electric field. Its key feature is a special shape of the extraction electrodes providing a higher rate of ion acceleration. It is applicable to any ion source type and could be used to improve performance of a wide range of plasma devices. The proof of concept was carried out using a high-current electron-cyclotron resonance ion source SMIS 37. High efficiency of the new extraction and proton beam formation with a record current density of up to $1.1~\text{A}~\text{cm}^{-2}$ was demonstrated.
\end{abstract}

%
\vspace{2pc}

\maketitle
 
\ioptwocol

\section{Introduction}

The difficulties in the high-quality and high-intensity ion beams formation are due to the significant space charge, which is the most influential at the initial stage of acceleration, when the ion energy is still low. In \cite{vybin2020} the authors of this article proposed and numerically substantiated the original approach to a significant expansion of the possibilities for the formation of ion beams, associated with the use of shaped electrodes that provide a highly nonuniform distribution of the electric field in the interelectrode gap. The shape of the electrodes is chosen to achieve the major drop in the accelerating potential being located close to the plasma electrode, thereby increasing the rate of particle acceleration in the region most sensitive to self-action (the region with a low ion velocity, thus, high space charge). A feature of the proposed system is that it increases the electric field only in the area where it is necessary (in contrast to traditional systems in which the field increases throughout the accelerating gap).

The new approach of the extraction system design was proposed in \cite{vybin2020}. It is applicable to every ion beam source, no matter of its type. This work shows an example of the new geometry usage at a particular ion beam source - SMIS 37. The experiment is carried out to prove the concept and confirm conclusions which are made in \cite{vybin2020}. 

\begin{figure*}[t]
\centering
\includegraphics{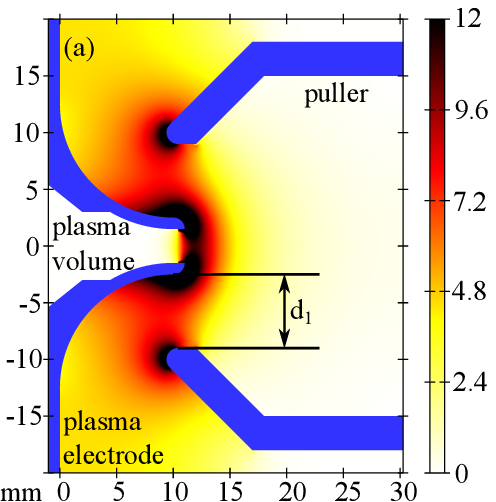}
\includegraphics{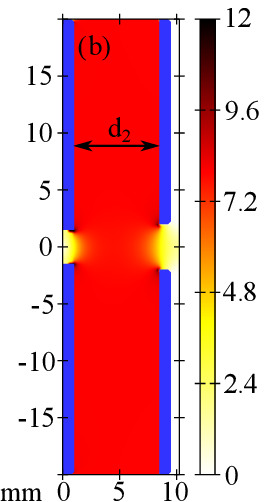}
\includegraphics{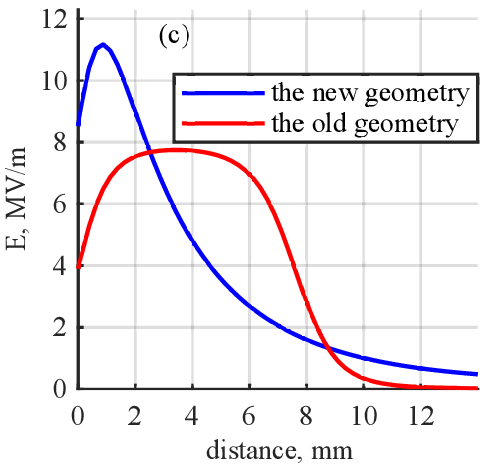}
\caption{\label{fig:fig1} The comparison of the flat extractor and the system with inhomogeneity of the electric field. The interelectrode distances ($d_1=d_2$) and bias voltages are equal. The significant increase of the electric field modulus near the plasma electrode tip is  demonstrated with colormaps (a, b) and distribution on the system axis (c).
}
\end{figure*}

The considered system with inhomogeneity of the electric field is shown in Fig.~\ref{fig:fig1}a. The increase of the puller hole diameter does not significantly affect the ion beam formation, but it reduces the ion flux on the puller and improves the extraction system reliability. For comparison, the flat extractor is also shown in the figure. A significant increase in the electric field is achieved in the new geometry at the same voltage. This is illustrated with colormaps of the electric field modulus (a, b), and graphs of the electric field distributions on the system axis (c).

\begin{figure*}[t]
\centering
    \includegraphics{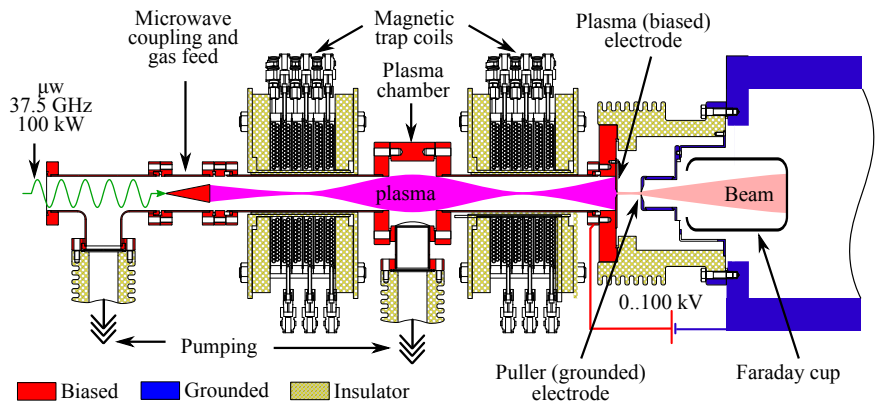}
    \caption{\label{fig:fig2} The SMIS 37 facility scheme.}
\end{figure*}
The use of the extraction system with highly inhomogeneous distribution of the electric field will make it possible to achieve significant progress in the characteristics of the generated ion beams. It allows for either an increase in the current density of the extracted beam at a constant accelerating voltage, or a decrease in the optimal extraction voltage in the case when the current density is determined by the plasma emissivity. An additional advantage of the new geometry is its weak sensitivity to the interelectrode distance, which makes it possible to effectively solve technical and engineering problems common for ion optics, in particular, those related to the electric breakdown resistance of the accelerating gap \cite{vybin2020}.

Improvement of the ion beam formation system makes it possible to achieve substantial progress in challenges, where the existing ion sources are not efficient enough. First, the increase of the total ion beam current without losses in its quality is feasible when the ion source with extra emission ability is used as an injector for modern accelerators. Second, the effective formation of a low energy ion beam is required for ion implantation \cite{ion_impl}. Third, the use of the new approach can be especially beneficial for the formation of focused ion beams (FIBs), since it allows one to significantly exceed the limitations on the current density associated with the Child-Langmuir law in the planar case. Ion sources for FIB are characterized by small apertures of electrodes in comparison with interelectrode distances \cite{fibs}. Decrease in the interelectrode distance is limited by the electrical breakdown of the interelectrode gap and by technological difficulties (accuracy of positioning the electrodes). This aspect ratio is far from optimal in the case of the flat geometry, and the use of the proposed geometry greatly simplifies the solution of the problem.

It should be noted that the new approach to the formation of an ion beam is also applicable to a multicomponent plasma. In particular, when forming a beam of negative ions, it is possible to decrease the power of the co-extracted electron beam \cite{3D_Ch-L}.
The use of the new approach makes it possible to unleash the potential of those facilities, where the plasma emission ability exceeds the capability of the extraction system \cite{SMIS_overview}.

This article is devoted to the experimental verification of the proposals formed in \cite{vybin2020}. The SMIS 37 \cite{SMIS_overview} pulsed facility was chosen as an experimental base. It is a high-current ion source based on plasma of an electron cyclotron resonance (ECR) discharge, sustained by the radiation of a gyrotron with a frequency of $37.5~\text{GHz}$ and a power of up to $100~\text{kW}$ in a magnetic trap of a simple mirror configuration. The combination of high frequency and high microwave power allows for the formation of dense plasma fluxes (ion flux up to $10~\text{A}~\text{cm}^{-2}$) through the magnetic trap plugs, which are used to obtain ion beams with record brightness for ECR ion sources.

SMIS 37 is a promising ion source with the so-called quasi-gas-dynamic plasma confinement mode \cite{SMIS-GDS-MCI, SMIS-GDS-MCI-2}, so-called a gas-dynamic ECR ion source \cite{SMIS_overview}. To verify the efficiency of the proposed approach, we compared the parameters of the beams formed by the conventional flat-electrode extraction system with the proposed one.

Earlier, SMIS 37 demonstrated the possibility of generating beams with unique parameters: proton beams with a current of up to $500~\text{mA}$ at a current density of $800~\text{mA}~\text{cm}^{-2}$, a fraction of atomic ions of more than $95\%$, and a normalized rms emittance not exceeding $0.1~\pi~\text{mm}~\text{mrad}$ \cite{SMIS-Proton}; beams of multiply charged nitrogen and argon ions with current up to $200~\text{mA}$ and an average charge of $4-5$ \cite{SMIS-MCI-New}.

Despite the record level of previously obtained results, there are a number of applications, which require the achievement of even higher currents (and current densities) of ion beams, while keeping the emittance low. The requirements of modern accelerators for light and multiply charged ion injectors are constantly growing, and to meet those of projects such as ISIS II \cite{ISIS-II} and HIAF \cite{HIAF} further enhancement of beam characteristics is needed. The use of the new extraction system with strongly inhomogeneous electric field distribution might be an important step towards the successful creation of the required ion sources not only for the mentioned projects, but also for many others, as well as make it possible to significantly improve the performance of existing systems through an easy and low-cost upgrade.

\section{Experimental facility}
\begin{figure*}
\centering
   \includegraphics{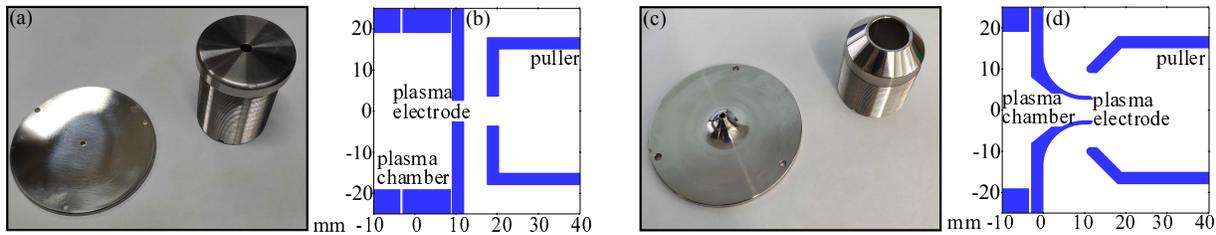}
    \caption{\label{fig:fig3} The electrode photos and the extraction system cross sections for the flat (a, b) and the “spherical” (c, d) geometry. The puller aperture in the spherical case is several times larger than in the flat one.}
\end{figure*}
The SMIS 37 facility scheme is shown in Fig.~\ref{fig:fig2}. Plasma is created using a pulsed microwave gyrotron radiation at a frequency of $37.5~\text{GHz}$, power up to $100~\text{kW}$ and a pulse duration of $1~\text{ms}$. Microwave beam is coupled to the discharge chamber using an electrodynamic system with an embedded gas feed line. Neutral gas is injected along the axis of the plasma chamber. The chamber inner diameter is $38~\text{mm}$. A simple magnetic trap with the mirror ratio of $R=5$ is used to create and confine the plasma. The distance between the plugs is $250~\text{mm}$. The magnetic field is created by the pulsed coils. The maximum magnetic field in the plug reaches $4~\text{T}$. The resonant value of the magnetic field for a frequency of $37.5~\text{GHz}$ is $1.34~\text{T}$. The pulse repetition period is about $20~\text{s}$.

With  hydrogen as a working gas, the plasma density reaches the level of $2\cdot10^{13}~\text{cm}^{-3}$, electron temperature -  $50~\text{eV}$ and ionization degree is close to $100\%$. This combination of temperature and density provides a high density of the plasma flux from the trap and corresponds to the quasi-gasdynamic confinement regime. The difference between this regime when compared to the  classical collision-less regime is that electrons fill the loss cone in the velocity space due to the high frequency of Coulomb scattering. The flux of electrons leaving the trap is limited by the ion flux, the maximum velocity of which is equal to the ion-sound one $V_s=\sqrt{T_e/M}$. The lifetime of electrons is determined by the mirror ratio, ion-sound velocity and trap length: $\tau_e = LR/V_s$. Substituting the typical values, one gets $\tau_e \sim 20~\mu\text{s}$. High plasma density and short lifetime provides an ion current density in the plug of more than $10~\text{A}~\text{cm}^{-2}$. To reduce the flux density to a level acceptable for beam formation, the extraction system is moved from the plug to the region of a weaker magnetic field. The distance from the plug of the magnetic trap to the plasma electrode is $120~\text{mm}$. The value of the magnetic field at the edge of the plasma electrode is about 10 times lower than in the plug.

\section{The beam extraction system}

The two-electrode system for the ion beam formation, consisting of a plasma electrode and a puller, was used. The maximum available extraction voltage on the facility is $100~\text{kV}$. In the experiment, the current of the ion beam passed through the formation system was measured using a Faraday cup (FC), the size of which ensures the interception of the entire beam.

Single-aperture two-electrode extraction systems are characterized by three parameters: the plasma electrode aperture $D_1$, the interelectrode distance $L$, and the puller electrode aperture $D_2$. 

The interelectrode distance is longer than plasma electrode and puller apertures for the majority of flat extraction systems. These extractors suffer from secondary electron emission in the case when the ion beam current density is too high and the beam significantly diverges in the interelectrode region. Therefore, the extraction system with significantly wider puller aperture was used (see Fig.~\ref{fig:fig3}).

In \cite{vybin2020} it was shown that the new approach allows for a significant increase in $D_2$ without a noticeable decrease in the quality of beam formation, while the flat extraction system efficiency suffers a lot from the  increase in the puller aperture. The diameter of the puller hole in the flat geometry in the experiment was significantly smaller than in the new one (see Fig.~\ref{fig:fig3}). The plasma electrode in the flat geometry has a considerable thickness (see Fig.~\ref{fig:fig3}b). This is necessary in order to ensure the same boundary conditions for the plasma flow in the considered cases.

\section{Experimental results}
In the general case, the dependence of the beam current on the Faraday cup on the extraction voltage $I_{FC}(U_{ext})$ has 2 regions with different modes of beam formation. At low extraction voltages, a space charge limited mode is realized, which is characterized by $I_{FC} \sim U_{ext}^{3/2}$ and corresponds to the Child-Langmuir law. At higher voltages, the plasma-density-limited mode (saturation mode) is realized, which is characterized by a constant current value.
An experimental comparison of the efficiency of the extraction systems was made for two extraction systems with 3 mm and 5 mm plasma electrode aperture. The experiments were carried out in modes with different plasma emissivity.

\subsection{Decrease of the optimal extraction voltage}

\begin{figure}
    \includegraphics{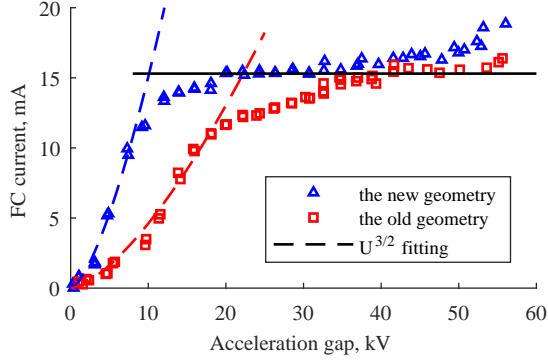}
    \centering
    \caption{\label{fig:fig4} The dependence of the Faraday cup current on the extraction voltage. The plasma electrode aperture is 3 mm.}
\end{figure}

Graphs of $I_{FC}(U_{ext})$ are shown for the 3 mm extraction hole case in Fig.~\ref{fig:fig4}. In this case, the plasma emissivity was moderate, which made it possible to achieve two extraction modes (firstly, the space charge limited mode and then the plasma density limited mode) at available accelerating voltages.
\begin{figure}[b]
\includegraphics{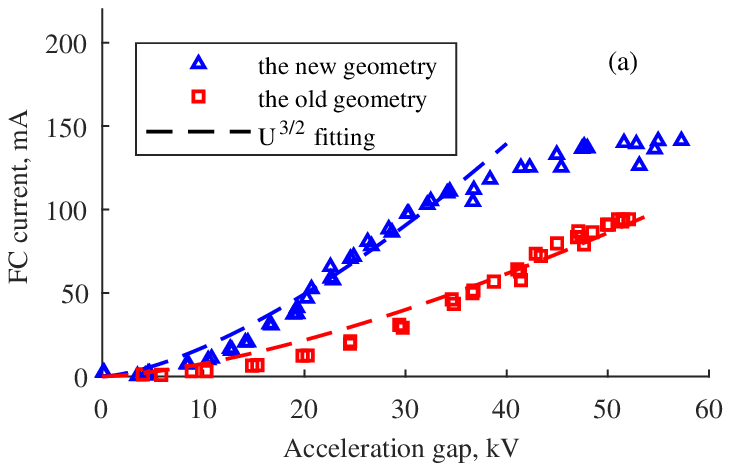}
\\
\includegraphics{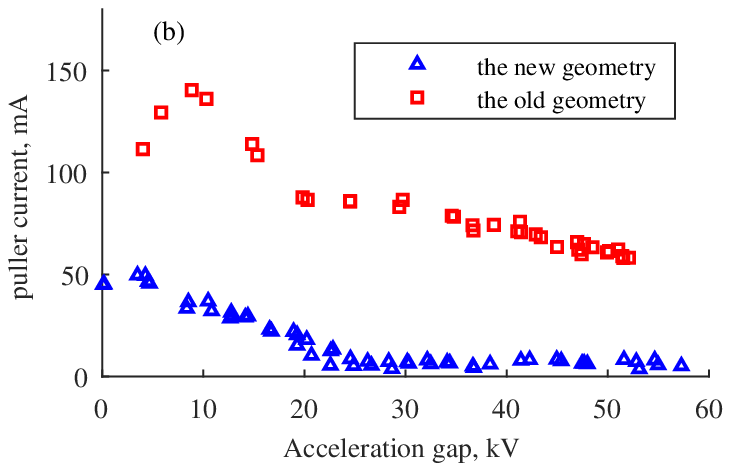}
   \centering
    \caption{\label{fig:fig5}
    The dependence of the Faraday cup (a) and puller (b) currents on the extraction voltage. The plasma electrode aperture is 5 mm.
}
\end{figure}
The use of the new ion beam formation system can significantly reduce the optimal extraction voltage. Consider the point of intersection of the approximation curves of the two extraction modes described above. For the flat case, the voltage of the characteristic point is $22~\text{kV}$ and for the new geometry - $10~\text{kV}$. Therefore, a significant decrease in the optimal extraction voltage (by a factor of 2) is shown when using the new geometry of electrodes.

\subsection{Reduction of the puller current}

In the following case the plasma emissivity was higher. For the flat case only the space charge limited regime was achieved. Graphs of the Faraday cup and puller currents dependencies on the extraction voltage are shown for the 5 mm extraction hole case in Fig.~\ref{fig:fig5}. It is clearly seen that in this case the flat system of electrodes did not allow for the utilization of the available plasma flow, while the new system ensured the saturation of the extracted beam current and its significantly higher value. The use of a tubular puller with a large aperture in the new geometry made it possible not only to increase the beam current, but also to significantly reduce the ion flux to the puller (see Fig.~\ref{fig:fig5}b). This effect was predicted in \cite{vybin2020} and it is a significant technological advantage of the developed extraction system.

\begin{figure}[b]
\centering
   \includegraphics{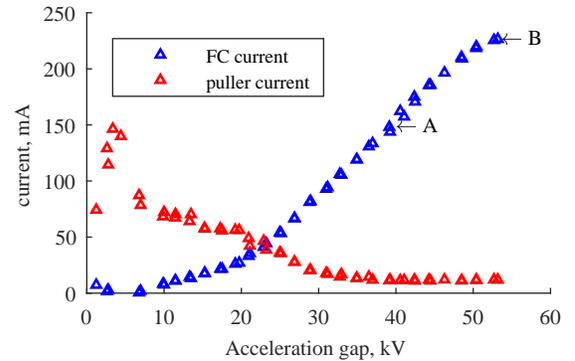}
   \centering
    \caption{\label{fig:fig6} The dependence of the Faraday cup and puller currents on the extraction voltage for the new geometry in the maximum current density mode. The plasma electrode aperture is 5 mm. Beam current waveforms in the points A and B are shown in figure~\ref{fig:fig7}.}
\end{figure}

Let us compare the efficiency of the new geometry for the given configurations. For this, we estimate the ratio of perveances $P$ of different extraction systems. The following quantity is called perveance
$$P=\frac{I_{beam}}{U_{ext}^{3/2}}.$$
The value of perveance indicates how significant the space charge effect is on the beam’s motion and is constant for the fixed geometry while operating in a space-charge-limited mode. The value of the perveance ratio for the 3 mm extraction hole case is $P_{new}/P_{old}=3.3$ and for the 5 mm case is $P_{new}/P_{old}=2.3$.

The increase in perveance ratio, thus, the efficiency of the new extraction system, becomes greater with the higher $L/D_1$ ratio. This fact confirms the conclusions drawn in \cite{vybin2020}.

\subsection{The record current density beam formation}
\begin{figure}
\centering
   \includegraphics{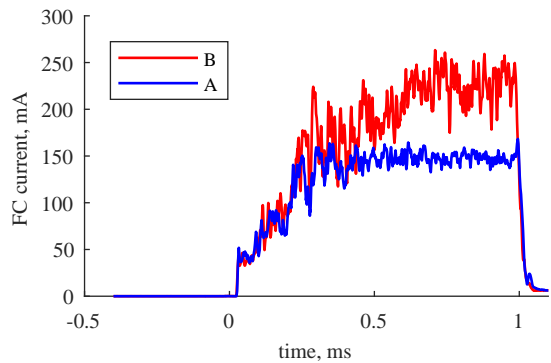}
    \caption{\label{fig:fig7} The Faraday cup current oscillograms at points A and B from figure~\ref{fig:fig6}. 
    }
\end{figure}
The most significant experiment which proves the advantages of the proposed extraction system with an inhomogeneous electric field distribution is the demonstration of the possibility of the ion beam formation with a record current density. The graph of $I_{FC}(U_{ext})$ is shown for the 5 mm extraction hole case in Fig.~\ref{fig:fig6}.

Starting from a voltage of about $35~\text{kV}$, the current to the puller reaches its minimum value. In this configuration, ion beams with currents from $150~\text{mA}$ (which corresponds to an average current density of $0.77~\text{A}~\text{cm}^{-2}$) at the extraction voltage of $40~\text{kV}$ to $225~\text{mA}$ (which corresponds to an average current density of $1.15~\text{A}~\text{cm}^{-2}$) at the voltage of $53~\text{kV}$ were obtained. Waveforms of the beam current for points A, B in Fig.~\ref{fig:fig6} are shown in Fig.~\ref{fig:fig7}.

\section{Conclusion}

The new extraction system geometry makes it possible to overcome the limitations on the ion beam current density, caused by the Child-Langmuir law for flat geometry. The assumptions (theoretically substantiated in \cite{vybin2020}) about the effectiveness of using the new geometry of the electrodes for obtaining high-current ion beams were successfully confirmed experimentally. It is shown that under the same conditions the optimal voltage required for high-quality beam formation using the system with inhomogeneity of the electric field is reduced by a factor of two when compared to the flat geometry. This result demonstrates the possibility of a significant increase in the beam current in systems with a fixed extraction voltage with a sufficient emissivity of the plasma. The relaxation of the requirements for the puller geometry not causing the deterioration of the beam quality was clearly demonstrated. An increase in the puller aperture, and even the use of the tubular puller, made it possible to minimize the contact of the puller electrode with the ion beam, eliminate secondary electrons emission from the puller and increase the electrical breakdown resistance of the interelectrode gap. During the tests of the new system, a proton beam with a record current density for any type of ECR ion sources at the level of $1.15~\text{A}~\text{cm}^{-2}$ was obtained.  
Experimental demonstration of the new extraction system efficiency makes it possible to proceed to the next stages of its implementation at facilities of various classes and purposes, sources of multiply charged ions, sources for ion implantation, sources of negative ions, etc.

\ack 
The work was supported by the project of the Russian Science Foundation Grant No. 21-19-00844.

\end{document}